\documentclass{elsart}

\usepackage{natbib}

\usepackage{amssymb}

\usepackage{graphicx}

\begin{document}

\begin{frontmatter}



\title{Efficient control of transient wave forms to prevent spreading depolarizations}


\author {M.\ A.\ Dahlem\corauthref{cor1},}
\author {F.\ M.\  Schneider,}
\author {and E.\ Sch\"oll}

\corauth[cor1]{dahlem@physik.tu-berlin.de}

\address{Institut f{\"u}r Theoretische Physik, Technische Universit{\"a}t Berlin,  Hardenbergstra{\ss}e 36, D-10623 Berlin, Germany}

\begin{abstract}
In various neurological disorders spatio-temporal excitation patterns
constitute examples of excitable behavior emerging from pathological
pathways. During migraine, seizure, and stroke an initially localized
pathological state can temporarily spread indicating a transition from
non-excitable to excitable behavior.  We investigate these transient
wave forms in the generic FitzHugh-Nagumo (FHN) system of excitable
media.  Our goal is to define an efficient control minimizing the
volume of invaded tissue.  The general point of such a therapeutic
optimization is how to apply control theory in the framework of
structures in differential geometry by regarding parameter plane $M$
of the FHN system as a differential manifold endowed with a metric.
We suggest to equip $M$ with a metric given by
pharmacokinetic-pharmacodynamic models of drug receptor interaction.
\end{abstract}

\begin{keyword}
nonlinear  dynamical system \sep excitability \sep control
\PACS 05.45.-a	
\end{keyword}

\end{frontmatter}



\section{Introduction}

Excitability is a central concept in neurophysiology both in health
and disease \citep{HIL97}.  Under some pathological conditions altered
excitability can spread through the tissue. Such transient wave forms
(TWF) occur for  example in migraine
with aura \citep{LAU94,AUR98,WEL05}, epilepsy
\citep{SOM84,GIO06}, and stroke
\citep{NED86,HOS94,FIS95,Hak98,Bac98}. In
migraine with aura seizure-like activity spreads slowly through parts
of the cortex. This is observed by symptomatic \citep{Las41} and
electrophysiological \citep{Had01} events.  Likewise,
epileptic seizures can have a localized onset and then grow in
intensity and start to spread. This usually leads to subsequent
generalized motor involvement commonly referred to as partial seizures
with secondary generalization. In some cases, however, the epileptiform
activity may induce changes to subcortical structures producing
clinical signs of general motor involvement that merely mimic a spread
\citep{Sch07c}.  Last but not least, during stroke a cortical
region that surrounds the infarct core and that initially suffers
functional injury can gradually and progressively fail and suffer
irreversible structural injury in untreated patients \citep{Bai97}.
In this condition the electrophysiological changes are persistent, but
there exist intermediate repetitive TWF and it was suggested
that 'therapy might [...] target the intermediate
forms of spreading depolarizations so as to protect the
penumbra [tissue surrounding the infarct core] against recruitment into the infarct core' \citep{DRE06}.
These three paradigmatic clinical manifestations of spreading
pathological states motivate efforts to understand how the spread of
such states arises and how it can be controlled.


%

The phenotypical manifestation of such spreading pathological states
is the phenomenon of spreading depression (SD) \citep{Lea44}.  SD is
the basis of migraine with aura. Essentially identical
electrophysiological features are associated with infarct expansion,
called peri-infarct depolarization (PID) \citep{May96,Str02,Fab06}.
The cascade of events that produce SD is related to seizures
\citep{Kag00}.  Despite certain differences, seizure events that begin
with epileptiform discharges can either terminate in SD, facilitate
the synchronization, or spread by a similar mechanism over a large
area with a velocity resembling that of SD \citep{Bur74,SOM84,Gor04}.
SD is seizure-like activity evolving as a slowly spreading non-or-all
type process. It is characterized by the feedback of ion currents that
change ion concentrations, which, in turn, influence the membrane
potential \citep{Kag00,Som01}. Shortly after its onset all neuronal
activity is depressed, hence its name. The name is misleading, because
SD can still be observed even when neuronal activity is depressed by
blocking the fast transient sodium current I$_{Na}$ \citep{Tob82}.  SD
emerges from an excitable pathway in neuronal tissue independently of
the normal neuronal activity. It was therefore suggested to categorize
SD and similar phenomena under the term {\it spreading depolarizations}
\citep{DRE06}.

There is ample evidence that SD belongs to the self-organization
processes due to the coupling of biochemical
reaction with diffusion \citep{Mar00a,Dah04a}.
Mathematical models of SD have been suggested
\citep{Tuc78,Reg94,Sha01,Kag00,Som01}, though
there is not yet  consent on the mechanism. We will use the
spatially extended FitzHugh-Nagumo (FHN) system as a generic model of
neuronal excitation patterns based on reaction-diffusion.  As a
neuronal model, it describes generic pattern formation properties not
limited to nerve impulse propagation along an axon, although it was
originally derived from the Hodgkin-Huxley model of action potentials
\citep{Hod52,Fit61,Nag62}.  The FHN system describes
also the spatial features of SD wave in neuronal tissue
\citep{Dah04b}.  Furthermore, the transition from non-excitable to
excitable media supporting traveling waves was well investigated in
the FHN model \citep{Win91,Hak99}.  It was suggested that the
spatio-temporal patterns in SD  occur at this transition
\citep{Dah04b}.  The regime in which this transition takes place
is also well investigated in chemical model systems in experiment and
theory, for a review see \citep{Mik06}.

The route to spreading depolarizations in a generic model is provided
by two independent pathways: one lowering the threshold of evoked
pathological activity, the other changing the time scale of
biochemical reaction rates, and hence their time scale ratio.  The two
pathways might offer new opportunities in developing optimal
therapy. Consider the case that a pathological condition is caused by
a shift along one path whereas therapeutic strategies are available
for both pathways. Suppose both strategies can be combined while each
has individual response rates and side effects.  What is an optimal
time efficient combined therapy stopping the spread while minimizing
side effects?  Since therapies can be combined, there is a two
dimensional manifold in which therapy takes place.  The strategy we
suggest is to equip this manifold with a metric that allows us to find
an effective combined therapy with minimal side effects.
Effectiveness is defined by finding a path to a sufficiently low
excitability where the tissue is not susceptible to spreading events,
whereas efficiency refers to side effects and time.

\section{Parameter space of the FHN system}

\begin{figure}[!b]
\centerline{\includegraphics[width=0.75\textwidth]{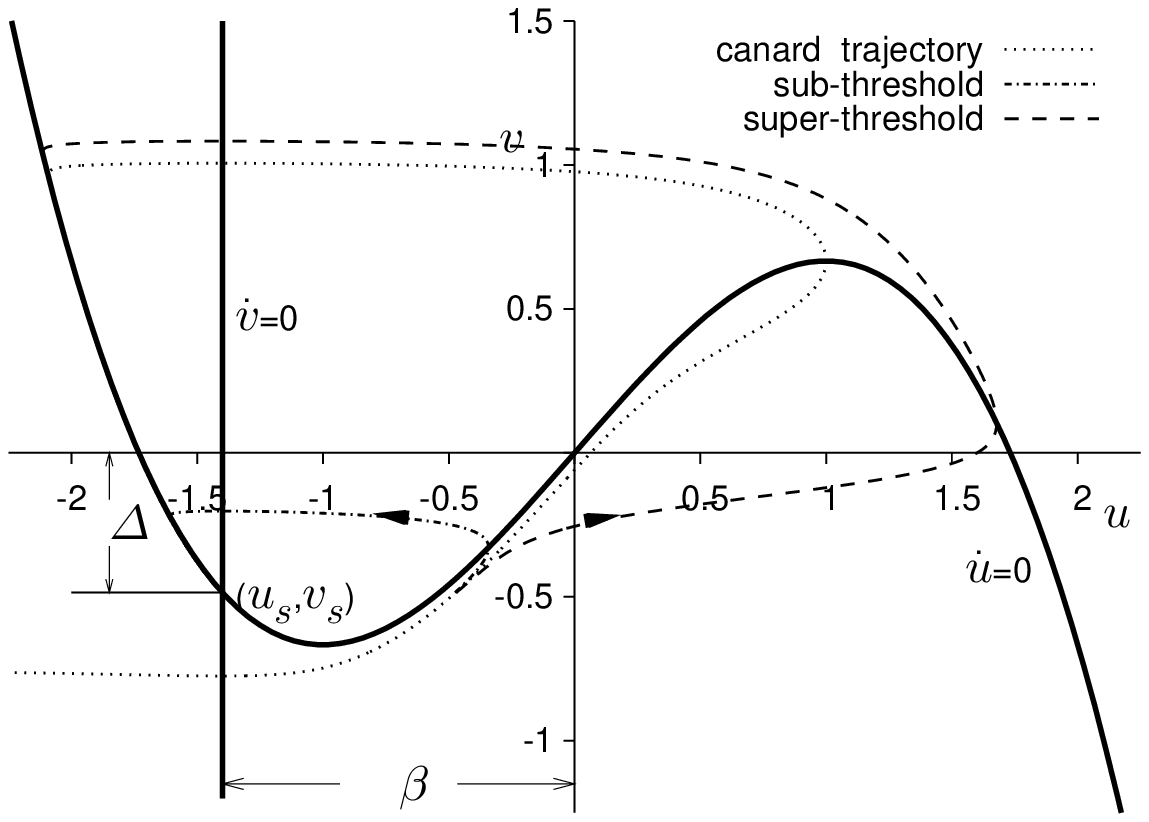}}

\caption{\label{fig:phasespace}
   The nullclines (bold) $\dot u=0$ and $\dot v=0$ in the phase space of the
   homogeneous FHN system with $\gamma=0,\beta=1.4$. Their intersection at
   $(u_s,v_s)$ is a stable fixed point. Three trajectories are drawn
   for  $\epsilon=0.04$: one canard trajectory (dotted), passing through
   the maximum of the nullisocline $\dot u=0$, and two
   trajectories starting at $v=v_s$  nearby but on
   opposite sides of the canard trajectory. They diverge
   sharply,
   producing threshold behavior: (dashed) = super-threshold  and
   (dashed-dotted) = sub-threshold stimulation.  The parameter
   $\beta$ correlates with the
   threshold size, while $\Delta$ is in a certain range inversely related and therefore
   correlates with the excitability of the system (see text).}
\end{figure}

We assume that a standard activator-inhibitor scheme leads to the
observed propagation phenomena during SD.  The activator and
inhibitor variables, $u$ and $v$, are coupled by their kinetic
reaction rates $f(u,v)$ and $g(u,v)$, respectively, and can diffuse in the
medium. The equations are

\begin{eqnarray}
\label{eq:fn}
\frac{\partial u}{\partial t} &=& f(u,v) + D_u \frac{\partial^2
   u}{\partial x^2}\\ \nonumber
\frac{\partial v}{\partial t} &=& \epsilon g(u,v) + D_v \frac{\partial^2
   v}{\partial x^2}.
\end{eqnarray}
Diffusion is represented by diffusion coefficients $D_u$ and $D_v$.
By re-scaling space, the ratio $\delta=\frac{D_v}{D_u}$ of the
diffusion coefficients can be introduced replacing the parameters $D_u$ and $D_v$. 
The parameter $\epsilon$ is the time scale
ratio of inhibitor and activator variables.  The reaction rates
$f(u,v)$ and $g(u,v)$ may possibly be
  derived from a more complex  model of SD, e.g., the one from \cite{Kag00}, by
lumping together all activator variables, such as inward currents and
extracellular potassium concentration $[K^+]_o$ into a single
activator variable and their combined kinetics into a reaction
rate $f(\cdot,\cdot)$. Likewise, a single inhibitor variable could
be related to recovery processes, such as effective regulation of
$[K^+]_o$ by the neuron's $Na$-$K$ ion pump and the glia-endothelial
system. This will be an important task of future investigations. We
aim to describe universal features of reaction-diffusion coupling that
lead to the onset of spreading pathological states and do not specify
the variables $u$ and $v$ which underly these characteristics other
than that they play the roles of activator and inhibitor,
respectively.  Their kinetic functions $f(u,v)$ and $g(u,v)$ are given
by the FHN system

\begin{eqnarray}
f(u,v) &=&   u -\frac{1}{3} u^3 -v\\ \nonumber
g(u,v)&=& u+\beta - \gamma v
\end{eqnarray}
where $\beta$ and $\gamma$ determine the excitation threshold (Fig.\ \ref{fig:phasespace}).


We shall start by viewing the parameter plane of the FHN model as a
manifold $M$ and consider its geometric structure.  $M$ has four
dimensions ($\epsilon,\beta,\gamma,\delta$). In a certain regime the
parameters $\beta$ and $\gamma$ determine the threshold for a
non-or-all excitation process. Like $\epsilon$, their variation can
cause a bifurcation: the emergence of sustained travelling
waves. Instead of the four dimensions usually a two dimensional subset
is investigated to describe this bifurcation, for example the section
at $\gamma=0.5$  \citep{Win91}, or at $\gamma=0$   \citep{Hak99},
both with $\delta=0$.  Consider  the  subset at $\gamma=0$. A
particular FHN system is specified by a point $q$ of this subset. It can be
parameterized by the coordinate functions $\epsilon(q)$, i.\,e., the
time scale separation of $u$ and $v$, and the threshold parameters
$\beta(q)$.
As an alternative coordinate function for $\beta(q)$

\begin{equation}
\label{eq:ct}
\Delta(p)= (\beta-\frac{1}{3}\beta^3)
\end{equation}
can be chosen \citep{Hak99}.  While $\beta$ is a measure of the
threshold, $\Delta$ is related to a measure of excitability, because
it is equal to the inhibitor concentration in the steady state (Fig.\
\ref{fig:phasespace}).  This is rather a convention than a definition
of excitability.  To be more general, we shall only assume that
excitability $E$ is a $C^\infty$-function $ E: U \rightarrow
\mathbb{R}$ in a subset $U$ of $M$. Further properties of this
function will be defined later. Firstly we want to note that there
obviously exists a whole set of coordinate systems $\mathcal{A}$ for
$M$ with coordinate transformations like Eq.\ (\ref{eq:ct}) being
$C^\infty$ diffeomorphisms. Naturally, $M$ can be identified as a
$C^\infty$ differential manifold. For a given subset $U$ of dimension
$n$ we can choose the coordinate system $\xi = [\xi^1, \dots, \xi^n]
=[\xi^i] \in \mathcal{A}$, with $i \in \{1,2, \dots, n\}$ that seems
to suit best the purpose of study. For example, in the section with
$n\!=\!2$ and $\xi^1\!=\!\epsilon$ the parameter $\xi^2$ can be chosen
either as $\beta$ or as $\Delta$.  We will consider the hypersection
at $\delta=0$, and $\gamma=0$ and use various coordinate systems in
this submanifold.

\section{Transient Wave Forms (TWF)}

\begin{figure}[!b]
\centerline{\includegraphics[width=1.0\textwidth]{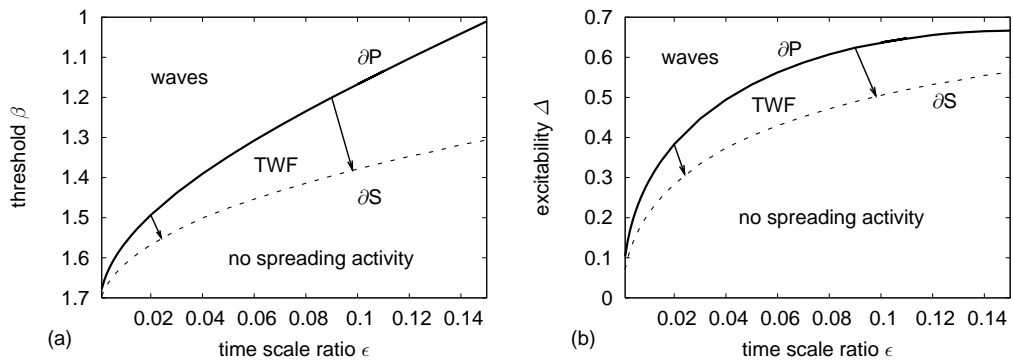}}
\caption{\label{fig:parameterSpace} 
Parameter space of the FHN
system ($\gamma=0,\delta=0$). Besides the parameter $\epsilon$, a {\it threshold parameter}
$\beta$ and an {\it excitability parameter} $\Delta$ is used in (a) and
(b), respectively, to span the parameter space (see Fig.\ \ref{fig:phasespace}).
Three
regimes exist defined by the spatio-temporal patterns that  occur:
waves, transient wave forms (TWF) and no spreading activity. 
The arrows mark two paths which are
perpendicular to the border $\partial S$ in (a) but not in (b).}
\end{figure}

Excitability is an emergent property of active media. It arises when a
critical parameter value is crossed above which the medium is
susceptible for sustained propagating excitation patterns
\citep{Mik91,Win91,Hak99,MIH02}.  In a 1D medium this border in $M$ is
the propagation boundary $\partial P$. It is obtained by finding
solutions in a co-moving frame in the form
$u(x,t)=U(y),\,\,v(x,t)=V(y),\,\,z=x+ct$, where $c$ is an additional
parameter of the unknown wave propagation speed.  Travelling wave
solutions in the form of a single pulse are thus equivalent to the existence of a homoclinic orbit
satisfying a system of ordinary differential equations \cite[p.224
ff]{KUZ95}.  The transition marks a bifurcation of codimension
one. Therefore $\partial P$ is a hypersurface in $M$ separating the
regime supporting travelling waves from the non-excitable one.  In a
two-dimensional parameter space, spanned by $\epsilon$, and $\beta$ or
$\Delta$, $\partial P$ is a curve (Fig.\
\ref{fig:parameterSpace}). Below $\partial P$ any confined
perturbation of arbitrary profile decays eventually.  Above $\partial
P$ some wave profiles are stable and travel with constant velocity.

Consider the region {\em close} to $\partial P$, which is referred to
as sub-excitable. There transient wave forms (TWF) exist though sustained waves may
not. Note first that in the absence of a metric in $M$ closeness to
$\partial P$ can only be defined with reference to other borders by
which $M$ is further subdivided.  The adjacent bifurcation curve above
$\partial P$ is $\partial R$. It is only defined in systems with
more than one spatial dimension. In these systems $\partial R$ is the
border above which open wave fronts will not disappear because each
open end curls in to form a spiral. Spiral waves rotate and re-enter
multiple times the invaded tissue. Below $\partial R$, but still above
$\partial P$, TWF exist depending on the initial size of the
excitation. In 2D such waves occur in the shape of particle-like waves
\citep{MIH02}, which can be controlled by excitability gradients
\citep{SAK02}. In this study, we investigate the region below $\partial
P$ where unstable TWF exist in the FHN systems with one spatial
dimension.

\begin{figure}[!tpb]
\centerline{\includegraphics[width=\textwidth]{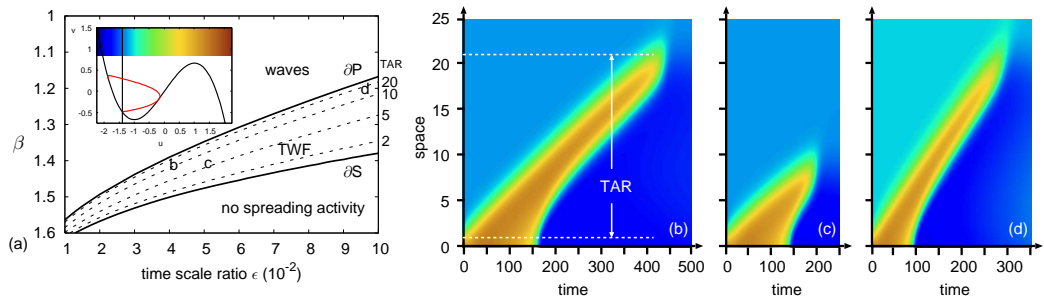}}
\caption{\label{fig:tar} 
(a) Parameter space with tissue-at-risk(TAR)-isolines
   (dashed). The extent a TWF spreads is defined from the stimulation
   border to the location where the maximum activator concentration $u$
   lies above the nullisocline $\dot u=0$ (inset). The red curve in the inset is
   the projection of a TWF from the infinite phase space of the space-dependent FHN system
   into the one of the homogeneous system for $\beta=1.4, \epsilon=0.04$. The
   projection is taken at the moment which we defined as the collapse of the TWF.
   (b)-(d) Space-time plots of the transient wave forms (TWF)
   following a stimulation (increase of $u$ by 2 for $0<x<1$, starting from the fixed point).
    Parameters: (b) $\epsilon=0.04$, $\beta=1.4$, (c) $\epsilon=0.05, \beta=1.4$,
    (d) $\epsilon=0.095, \beta=1.185$. $D_u=1$ and $\delta=0$ for all parts. The color code denotes the
    activator $u$ as defined in the inset of panel (a).}
\end{figure}

To further subdivide $M$ in the regime below $\partial P$, we suggest
to take the distance a TWF spreads (Fig.~\ref{fig:tar}b) to define
isolines.  This distance defines the volume of tissue at risk (TAR)
referring to the risk of transient neurological symptoms or even of
permanent damage (PID case) when cortical tissue is invaded following
a local stimulation.  At $\partial P$ stable traveling waves
exist. They can invade the whole tissue. Therefore the TAR value is
infinite at $\partial P$. Below $\partial P$ the value of TAR is
finite and rapidly decreases to zero when one moves the FHN system
away from $\partial P$ in parameter space (Fig.~\ref{fig:tar}b-c).  We found the
existence of a new boundary $\partial S$ as the TAR value approaches
zero. At this boundary any stimulation profile collapses into the
steady state without broadening, that is, without affecting
surrounding tissue.  $\partial S$ is therefore contrary to $\partial
P$ at which the TAR value is infinite.  The region between $\partial
S$ and $\partial P$ is the region where  TWF exist in a 1D system. This
region defines a subset $U$ of $M$. The general point of a therapeutic
control strategy is to leave $U$ by crossing $\partial S$.

\section{Efficient control of transient wave forms}

So far excitability $E$ has not been defined. We assume $E$ to be a
$C^\infty$-function in $U$. It is reasonable to assume that $E$ is
constant on the two borders, because these are  bifurcation lines at
which TAR is constant. Furthermore, since a change in TAR is
suggestive of a change in excitability, we propose both to be
linearly related.  

Our goal is to control a path $\Lambda$ in $U$ along which
excitability, and with it the volume of TAR, is efficiently
diminished. Let the path $\Lambda$ start in $U$ at $q$ where the
neuronal tissue temporarily supports the spread of excitation. Let the
path end on $\partial S$. Firstly, we assume sufficient knowledge of
control characteristics, e.\,g., by pharmacokinetic and
pharmacodynamic means, as described in the next section. We will show
in the next section that this provides a metric on $U$. In this
section we will just assume $U$ to be a differential manifold with a
metric. Furthermore, we assume (i) that our control method allows us
to choose any path $\Lambda:I \rightarrow U$ parameterized by some
interval $I \subset \mathbb{R}$, (ii), as already stated, that $E$ is
a $C^\infty$-function in a subset $U$, which includes $\Lambda$, and
(iii) that $ E(q) > E(\partial S)$ holds.

Which path should we take, if we want to reduce excitability by going from
$q$ to $\partial S$ by deliberate control?  The efficiency of successful
control
  dragging the system into the target state $p$ on $\partial S$ should be given by
some optimization criterion.  When a metric is given two paths are
privileged: the shortest path $\Lambda_s$ between $q$ and $\partial S$, and
also the one that minimizes $E$ by gradient descent. The latter path
$\Lambda_g$ implies a metric because covectors like $\partial
E/\partial \xi^j$ (representing a gradient) and contravariant vectors
(represented by tangent vectors to a path $\Lambda$) are
unrelated objects of different kinds.  Only a metric tensor $g_{ij}$ defines
the gradient as a tangent vector (using summation convention)

\begin{equation}
\label{eq:gd}
(\mbox{grad} E)^i = g^{ij}\frac{\partial  E}{\partial \xi^j}.
\end{equation}
Therefore, only when a metric is given on $U$, we can apply some
  optimization criterion for the efficient therapeutic path.


\section{Metric tensor on $U$}

As efficient control, we proposed in the last section a method based
on a metric structure in $U$. Therefore, a metric structure is needed
to optimize control. Hence it is natural in this context to endow $U$
with a metric that is derived by some sort of cost function of the
control method.  We introduce a standard pharmacokinetic and
pharmacodynamic scheme to illustrate this concept.  Let $\zeta^i$ be
the concentrations of drugs which regulate diverse functions in
populations of neurons.  For the sake of simplicity, we neglect the
details of pharmacokinetics as the discipline that describes dosage
regimes and the time-course of $\zeta^i$ in the body by absorption,
distribution, metabolism, and excretation.  We assume the drugs can be
constantly administered and their rate of administrations equals their
rate of metabolism and excretion.  Thus $\zeta^i$ is immediately in
its steady state value. Furthermore, we assume that $\zeta^i$ follows
linear pharmacokinetics. In this situation the steady state of
$\zeta^i$ changes proportionally according to dose.

\begin{figure}[!tpb]
\centerline{\includegraphics[width=\textwidth]{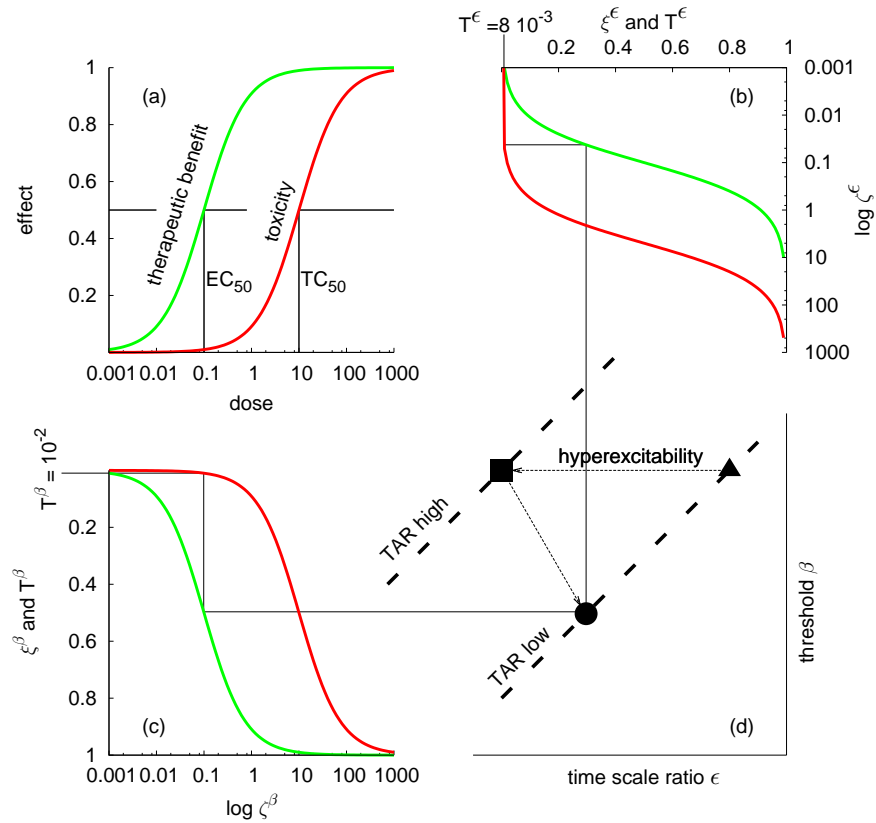}}
\caption{\label{fig:metric} (a) Dose response curves for an effective
dose and a toxic dose with the typical sigmoidal shape caused by the
logarithmic scaling. (b),(c) Same as in (a), but with rotated and inverted axes
for two drugs $\zeta^{\epsilon}$ acting on $\epsilon$ and 
$\zeta^{\beta}$ acting on $\beta$. 
(d) Schematic parameter space with two TAR-isolines (TAR high and TAR low). 
A route to hyperexcitability
can be caused by a shift in $\epsilon$ (from $\blacktriangle$
to  $\blacksquare$). The optimum path from the hyperexcitable state
($\blacksquare$) towards the isoline of the physiological state 
(TAR low) using the two dose and toxic response curves
in (b) and (c) leads to ({\Large \textbullet}). }
\end{figure}

The relation between drug dose and response is usually modeled as a
hyperbolic function assuming a simple drug receptor
interaction. Suppose the response to $\zeta^i$ is 
\begin{equation}
\label{eq:doseResponseRelation}
\xi^i = r^i\left(\frac{\xi^{i}_{max}  \zeta^i}{EC^i_{50} + \zeta^i}\right),
\end{equation}
where the  $\xi^i$ define a new coordinate system $\xi=[\xi^1, \dots,
\xi^n]$ in $U$.
In this equation, $r^i$ denotes transducer functions that represent
the response of the FHN system to the drug $\zeta^i$.  For the sake of
simplicity we use as transducer functions $r^i$ the identity. 
$EC^i_{50}$  are the {\it effective doses 50}, i.e., doses at which 50\% of
the maximal responses $\xi^{i}_{max}$ are achieved (Fig. \ref{fig:metric}a). $\xi^{i}_{max}$ is the
asymptotic value of $\xi^i$ 
for large concentrations.

In analogy with Eq.\ (\ref{eq:ct}), where we have introduced the new
coordinate $\Delta$, we have thus introduced the coordinate system
$\zeta\in \mathcal{A}$ as new control parameters of the FHN system.  A
new coordinate system with a known metric is needed as a reference
system from which the metric in $U$ is obtained by coordinate
transformation. 
Let us denote the components of the metric tensor of $U$ in the
coordinate system $\zeta$ as $g_{ij}$. Above we  have
 introduced the new coordinate system $\xi \in \mathcal{A}$ of the
response variables. The
components of the metric tensor in this  coordinate system ($\xi$) are then
\begin{equation}
\label{eq:pbm_co}
\tilde g_{\alpha\beta}=\frac{\partial \zeta^i}{\partial  \xi^\alpha} g_{ij}
\frac{\partial \zeta^j}{\partial  \xi^\beta}. 
\end{equation}
Likewise, we can calculate the unknown components $g_{ij}$ from
$\tilde g_{ij}$. The coordinate system $\zeta$ is a privileged
reference system for $U$ by selecting $\zeta^i$. However, this does
not imply that it is Cartesian ($\tilde g_{ij}=\delta_{ij}$,
$\delta_{ij}$ being the Kronecker delta tensor representing the
Euclidean metric in Cartesian coordinates), as is readily seen since the choice of
concentration units is arbitrary.  For example, commonly the logarithm
of the concentration $\zeta^i$ is plotted on the abscissa, as shown in
Fig. \ref{fig:metric}a. Hence the question is whether a structure of
the pharmacodynamic scheme can be natural choice treated as a metric
structure in differential geometry defining $\tilde g_{ij}$.

Suppose that $\zeta^i$ or one of their metabolites
have toxic side effects.  Their dose response curve follows the same
relation as Eq.\ (\ref{eq:doseResponseRelation}), although shifted to
the right on the dose axis by a higher {\it toxic dose 50} ($TC_{50}$)
\begin{equation}
\label{eq:dtr}
T^i = t^i\left(\frac{T^{i}_{max}  \zeta^i}{TC^i_{50} + \zeta^i}\right).
\end{equation}
$t^i$ denotes  transducer functions that represent the
  response of the toxic system.
Again for the sake of simplicity we use as transducer functions
$t^i$ the identity function. In analogy to Eq.\ (\ref{eq:pbm_co})
we obtain
\begin{equation}
\label{eq:pbm_co2}
\tilde g_{\alpha\beta}=
\frac{\partial \zeta^i}{\partial  \xi^\alpha}
\frac{\partial T^k}{\partial  \zeta^i}
\delta_{kl}
\frac{\partial T^l}{\partial  \zeta^j}
\frac{\partial \zeta^j}{\partial  \xi^\beta}, 
\end{equation}
assuming $T=[T^1, \dots, T^n]$ builds a Cartesian coordinate system of
the costs with the Euclidean metric $\delta_{kl}$. The association of
the cost coordinate system ($T$) with the Cartesian one is a naturally
choice, since the efficiency of control is--by definition--measured in
these costs.

We end this section with one example that to some extent corresponds
to a subtype of migraine with aura. For this subtype a novel pathogenic
genetic mutation was found.  Therefore, at least in principle, the
route to hyperexcitability was discovered on a molecular level. A recent
migraine study by \cite{VAN07} showed a  mutation in the gene
coding the sodium channel that leads to a slowed-down inactivation and
a two-fold faster recovery from inactivation.  In our simplified FHN
system this is modeled by slowed inhibition, i.\,e., lower $\epsilon$
values. As a consequence, the physiological state ($\blacktriangle$,
Fig.~\ref{fig:metric}d) of the cortex is shifted into hyperexcitable
state ($\blacksquare$). By definition, the physiological state
($\blacktriangle$) is on a lower TAR-level than the hyperexcitable
state ($\blacksquare$). The therapeutic goal would be to shift the
cortical state back to low a TAR-level, at which the pathological
activity, i.\,e., the neurological symptoms, do not spread.

To begin with, we need to specify dose response and toxicity curves.
They are chosen arbitrarily to complete the example but without direct
relation to the above mentioned subtype of migraine. Suppose that we
have a drug with concentration $\zeta^1$ acting exclusively on
$\epsilon$ (e.\,g., $\xi^1\equiv\epsilon$). Its dose response graph is
shown in Fig.~\ref{fig:metric}b. It is similar to the one in
Fig.~\ref{fig:metric}a but rotated such that the response axis
corresponds to the $\epsilon$-axis.  Let another drug $\zeta^2$ act
exclusively on $\beta$ (e.\,g., $\xi^2\equiv\beta$). Its dose response
graph is shown in Fig.~\ref{fig:metric}c. It is mirrored (from
Fig.~\ref{fig:metric}a) such that the response axis corresponds to the
$\beta$-axis.  For the sake of readability we replace the set $\{1,
2\}$ of indices $i$ by $\{\epsilon, \beta\}$ and use $\zeta^i$ for
both the drug concentration and the drug name. Let the effective doses
50 $EC^i_{50}$ be the same for both (without loss of generality \ we
choose $EC^i_{50}=0.1$), and let the toxic doses 50 $TC^i_{50}$
differ. Usually $TC^i_{50}$ is several times higher then the effective
dose. We choose $TC^\epsilon_{50}=5$ and $TC^\beta_{50}=10$. The drug
$\zeta^\beta$ has therefore a relatively higher therapeutic index
$TI^i=TC^i_{50}/EC^i_{50}$ than $\zeta^\epsilon$.  Furthermore, we
assume that both drugs have a sufficient efficacy, that is, they are
each potent enough to drag the FHN system back to it physiological
excitability defined by the TAR-isoline (Fig.\ \ref{fig:metric}d).
The maximal response rates ($\xi^i_{max}$) for both are chosen such
that already 80\% of $\xi^i_{max}$ is effective.  Likewise and again
without loss of generality we fix $T^i_{max}$ at $\xi^i_{max}$.

Given the example described above, we can calculate the cost for
individual and combined therapy.  If exclusively drug $\zeta^\beta$ is
administered to control excitability, the FHN system moves in
parameter space along the $\beta$-axis (downwards in
Fig.~\ref{fig:metric}d) until the physiological TAR level (TAR low) is
reached.  The toxicity imposed by the therapeutic benefit in this case
is 3.85\% of the maximal toxic level of $T^\beta$.  An exclusive
administration of drug $\zeta^\epsilon$ (which has a only half the
therapeutic index $TI^\epsilon$ compared to $TI^\beta$) cost 7.4\% of
the maximal toxic level $T^\epsilon$.  This corresponds to a shift
from point marked with $\blacksquare$ to the point marked with
$\blacktriangle$ in Fig.~\ref{fig:metric}d.  At last, we consider how
the two drugs must be administered in a combined therapy.  An optimal
combined therapy reaches the physiological TAR level at 1.84\% of the
maximal combined toxic level $T_{tot}=T^\epsilon+T^\beta$ (see the two
black lines starting at ({\Large \textbullet}) and terminating at the
both effect axes in Fig.\ \ref{fig:metric}b and d). Due to the
nonlinearity in the response and toxcity curves the total cost of the
combined theraby is much less than the average of the individual
therapies.

\section{Discussion}

The phenotypical scheme of cortical spreading depression describes transient
waves of massive depolarization of neurons and astrocytes. Such a {\em
spreading depolarization} \citep{DRE06} is associated with migraine,
epilepsy and stroke. Their etiologies are mainly discussed in the
context of hyperexcitability and disorders known as channelopathies,
that is, diseases caused by a mutation in gene coding for ion
channels.  The challenge is to bridge the gap between the molecular
level of the cause and macroscopic {\em tissue} level of the effects.

We suggest that there are at least two independed routes towards a
hyperexcitable state that supports transient wave forms in cortical tissue: one
changing the ratio of kinetic rates and one lowering the
threshold.  While the latter route changes the nullclines, the
former changes only the trajectories in the phase space. Consequently,
there are also two routes out of the hyperexcitable regime. This is of
particular interest when a critical therapeutic time window exits in
which the volume of affected tissue is largely increased, as for example
in peri-infarct depolarization. Then the therapeutic aim would be first to prevent tissue loss
within the given therapeutic time window and re-establish the physiological value
later. In Fig.\ \ref{fig:metric}d this would correspond to a path
from the hyperexcitable state ($\blacksquare$) directly to a state of
low tissue at risk ({\Large \textbullet}) and from there back to the
physiological state ($\blacktriangle$).

We have described efficient control of excitability by a simple
pharmacodynamic model.  In general, the study of beneficial effects of
independent pathways is complicated by numerous interactions between
pharmacokinetics, pharmacodynamics, and homeostatic factors and by
individual variability. For example, the introduction of antagonistic
behavior between $\zeta^\epsilon$ and $\zeta^\beta$ will complicate
the geometrical structure of $U$. In general the coordinate system of
the costs $[T^1, \dots, T^n]$ will not build a Cartesian coordinate
system. However, if we have a mapping from general costs and effects
defined by Eqs.\ (\ref{eq:doseResponseRelation}) and (\ref{eq:dtr}) we
still can infer a metric structure of the parameter space of
FHN.

In conclusion, as a first step to implement the proposed control
strategy accurate estimates given by pharmacokinetic-pharmacodynamic
models of drug activity are needed for the regime of transient wave
forms. We want to stress this point, namely that this estimates must
be obtained in the regime of transient wave forms, that is, a regime
close to a bifurcation. Attempts to obtain estimates of
pharmacokinetic-pharmacodynamic parameters in the regime far from this
bifurcation will probably lead to inaccurate results.  Then the
next step is to construct an appropriate metric on the differential
manifold of the parameter space in order to optimize the path from the
hyperexcitable state back to the physiological state.

\section*{Acknowledgments}

We are grateful to
  G. Bordyugov, and V. Zykov for fruitful discussions.
MAD would like to thank the Deutsche Forschungsgemeinschaft
for financial support (DA-602/1-1). This work was also supported within
the framework of Sfb 555.



\clearpage
\pagebreak

\clearpage
\pagebreak

\clearpage
\pagebreak

\end{document}